\documentclass[10pt,letterpaper]{article}
\usepackage[top=0.85in,left=2.1in,footskip=0.75in,marginparwidth=2in]{geometry}

\usepackage[utf8]{inputenc}

\usepackage{cite}
\usepackage{float}

\usepackage{amsmath}
\usepackage{amsmath}
\usepackage{amssymb}
\usepackage{nameref,hyperref}

\usepackage{lineno}
\usepackage{setspace}

\usepackage{microtype}
\DisableLigatures[f]{encoding = *, family = * }

\usepackage{changepage}

\usepackage[aboveskip=1pt,labelfont=bf,labelsep=period,singlelinecheck=off]{caption}

\makeatletter
\renewcommand{\@biblabel}[1]{\quad#1.}
\makeatother

\usepackage{lastpage,fancyhdr,graphicx}
\usepackage{epstopdf}
\fancyheadoffset[L]{2.25in}
\fancyfootoffset[L]{2.25in}

\usepackage{color}

\definecolor{Gray}{gray}{.25}

\usepackage{graphicx}

\usepackage{sidecap}

\usepackage{wrapfig}
\usepackage[pscoord]{eso-pic}
\usepackage[fulladjust]{marginnote}
\reversemarginpar

\begin{document}

\begin{flushleft}
{\Large
\textbf\newline{From temporal network data to the dynamics of social relationships}
}
\newline
\\
Valeria Gelardi\textsuperscript{a,b},
Didier Le Bail\textsuperscript{a},
Alain Barrat\textsuperscript{a,c},
Nicolas Claidiere\textsuperscript{b,d}
\\
\bigskip
\bf{a} {Aix Marseille Univ, Universit\'e de Toulon, CNRS, CPT, Marseille, France}
\\
\bf{b} {Aix Marseille Univ, CNRS, LPC, FED3C, Marseille, France}
\\
\bf{c} {Tokyo Tech World Research Hub Initiative (WRHI), Tokyo Institute of Technology, Tokyo, Japan}
\\
\bf{d} {Station de Primatologie-Celphedia, CNRS UAR846, Rousset, France}
\bigskip

\end{flushleft}

\section*{Abstract}
Networks are well-established representations of social systems, and temporal networks are widely used to study their dynamics. However, going from temporal network data, i.e., a stream of interactions between individuals, to a representation of the social group?s evolution, remains a challenge. Indeed, the temporal network at any specific time contains only the interactions taking place at that time and aggregating on successive time-windows also has important limitations. Here, we present a new framework to study the dynamic evolution of social networks based on the idea that social relationships are interdependent: as the time we can invest in social relationships is limited, reinforcing a relationship with someone is done at the expense of our relationships with others. We implement this interdependence in a parsimonious two-parameter model and apply it to several human and non-human primates? data sets to demonstrate that this model detects even small and short perturbations of the networks that cannot be detected using the standard technique of successive aggregated networks. Our model solves a long-standing problem by providing a simple and natural way to describe the dynamic evolution of social networks, with far-reaching consequences for the study of social networks and social evolution.

\newcommand{\cut}[1]{\textcolor{blue}{\sout{#1}}}
\newcommand{\add}[1]{\textcolor{black}{{#1}}}

\section*{Introduction}

Social relationships are created and maintained through interactions between individuals, which can last and be repeated over a variety of timescales. Social networks provide convenient representations for the resulting human and non-human animal social structures, where individuals are the nodes of the networks and links (ties) are summaries of their social interactions \cite{Granovetter:1973,Hinde:1976,Wasserman:1994,brent2011social}. Recently, the availability of temporally resolved data on interactions between individuals,
from various types of communication  \cite{Eckmann:2004,Kossinets:2006,onnela2007structure,Karsai_2011,miritello:2013timeallocation} to 
face-to-face interactions \cite{Cattuto:2010,salathe2010high,stopczynski2014measuring,Toth:2015role} has fueled the 
development of the field of temporal networks
\cite{holme_saramaki2012temporal,holme2015modern}, which replaces static ties by  
 information on the actual series of interactions on each tie,
\add{allowing} researchers to further the study of social networks. 
 For instance, aggregating temporal information over successive time windows has made it possible to follow the evolution of ties over larger timescales
\cite{saramaki2014persistence,Fournet:2014,gelardi2019detecting,aledavood2015digital}. Taking into account the temporal features of each tie during a certain time window can also shed light on their strength and persistence
\cite{navarro2017temporal,Urena-Carrion_Saramaki:2020}. Finally, researchers have identified temporal structures with no static equivalent
\cite{kovanen2011temporal,kobayashi2019the,galimberti2018mining} that reveal interesting patterns of relevance to the analysis of social phenomena or dynamic processes in a social group
\cite{kovanen2013temporal,ciaperoni2020relevance}.
 
Despite  \add{this progress}, moving from 
a stream of interactions within a group of individuals, represented by a temporal network, to a meaningful 
\add{quantification of the strength and}
evolution of their social relationships, remains a challenge. Indeed, 
the temporal network seen at any specific time $t$ contains by definition only the interactions taking place at $t$, \add{while the state of a relationship between two individuals at $t$
depends potentially on the whole history of their previous interactions, both mutual and with others.}
\add{Temporal aggregation over successive time windows is a commonly used approach to address this issue, but}
a number of properties of the networks obtained by temporal aggregation on successive windows depend on the window length and placement
\cite{sulo2010meaningful,krings2012effects,psorakis2012inferring,kivela2015estimating}. Aggregating over increasingly long time windows also 
averages out relevant temporal information \add{by treating in the same way old and recent interactions, and by not taking into account possible temporal correlations between
successive interactions, nor the impact of a single interaction on multiple ties. Moreover,} no single natural time scale
for aggregation can be defined, as relevant dynamics occur
on multiple timescales \cite{holme2013epidemiologically,saramaki2015seconds,Darst:2016,masuda2019detecting}.

Here, we \add{put} forward a new
systematic way to transform the stream of interactions between individuals into a continuously evolving representation of the social structure, i.e., a network with time-varying weights, \add{taking into account the temporal ordering 
of interactions in a non-trivial way}. 
The evolving weight $w_{ij}(t)$
of the tie between nodes $i$ and $j$ \add{represents a quantification of the strength} of their relationship at $t$.
\add{Moreover, our framework goes beyond the} 
 few such dynamic network models  proposed to date \cite{Ahmad:2018,zuo2019models,Jin2001StructureOG,Palla:2007}, \add{that are}
based on the idea that the weight of a tie between two individuals strengthens when they interact, and that in the absence of interaction, the tie's weight decays exponentially with time (the timescale of the decay is the model's parameter): these rules of evolution assume that the links between distinct pairs of individuals are independent, while the interdependence of social relationships is instead often well justified. 
For instance, 
\add{in the} complex social groups \add{formed by humans and other primates}
\cite{dunbar2007evolution, mitani2009male,Silk:2010}, 
 investing in a social relationship is a costly strategic decision that requires specific cognitive skills \cite{Cheney:1986} and the quality of an individual's social relationships depends on the time invested in them \cite{dunbar2020structure,dunbar2009time,borgeaud2021vervet}. 
\add{Thus}, the occurrence of a social interaction between two individuals not only reinforces their mutual relationship, but it also weakens the relationships they have with others: the time and energy spent to maintain the tie with an individual is taken from a finite interaction capacity and thus is time that is not spent with others. 
The framework that we put forward here \add{takes into account} this interdependence of social relationships 
to transform a stream of interactions into an evolving weighted network:
with each interaction between two individuals, the weight of their tie increases, while the weights of the ties they have with other individuals decrease. In
contrast to other recent temporal network representations \cite{Ahmad:2018,zuo2019models}, time itself 
is not explicit, and the weight of a tie remains unchanged if the corresponding individuals do not interact with anyone.
Our framework is therefore linked to the Elo rating method \cite{elo1978rating} used to rank chess players and analyze animal hierarchies: the dynamics of the system are determined by the pace of interactions between individuals, not by the absolute time between events. 

In the following, we define a parsimonious model for the evolution of social ties based on these concepts, with two parameters quantifying respectively the increase in the weight of a tie $i-j$ when an interaction occurs between $i$ and $j$, and its decrease when another interaction involving either $i$ or $j$ (but not both) takes place. We then show the relevance of the model by applying it to several data sets
describing interactions in groups of human and non-human primates and by using it to automatically detect 
naturally occurring changes in the groups' dynamics and artificially generated perturbations in the data.

\section*{Results}

\subsection*{Framework}

The concepts highlighted above can be translated in various ways to transform a stream of  interactions into evolving tie weights of an evolving \add{directed} network $G(t)$. The nodes of the 
network represent the individuals and the weight $w_{ij}(t)$ represents the strength of the social relationship \add{from $i$ to $j$} at time $t$. 
Here we consider a model depending on two parameters, $\alpha$ and $\beta$, with the following rules:
\begin{itemize}
    \item We start from an empty  network with uniform weights initialized to zero, i.e., $w_{ij}(0)=0 \quad \forall i,j$;
  \item
\add{At each time $t$, we denote by $E(t)$ the set of interacting ties at $t$. Then, for each tie $(i,j) \in E(t)$, 
the weights of the ties in which $i$ and $j$ are involved are updated according to}
\begin{eqnarray} \nonumber
w_{ij}(t^{+}) &=& w_{ij}(t^-) + \alpha(w_{max} - w_{ij}(t^-)) \\
w_{ji}(t^+) &=& w_{ji}(t^-) + \alpha(w_{max} - w_{ji}(t^-))
\label{equation:rule_update_+}    
\end{eqnarray}
and
\begin{eqnarray} \nonumber
w_{ik}(t^+) &=& (1 - \beta) w_{ik}(t^-)\ \ \forall k \ne j, \add{ (i,k) \notin E(t)} \\
w_{jk}(t^+) &=& (1 - \beta) w_{jk}(t^-)\ \  \forall k \ne i, \add{ (j,k) \notin E(t)} \ .
    \label{equation:rule_update_-}
\end{eqnarray}
\end{itemize}
\add{The weights of all ties interacting at $t$ thus increase according to \eqref{equation:rule_update_+}, while the weights 
of the neighbouring ties that do not interact at $t$ decrease according to \eqref{equation:rule_update_-}.}
\add{These rules of evolution can be applied to temporal network data expressed either in continuous 
time (i.e., an interaction between two individuals can occur
at any time) or in discrete time (when the data itself
has a finite temporal resolution): in the former case, 
 $t^-$ and $t^+$ stand respectively for the times immediately before and after the interaction; 
in the latter case, $t^-$ is simply replaced by $t$ and $t^+$ by $t+1$ in Eqs \eqref{equation:rule_update_+}-\eqref{equation:rule_update_-}.}
The parameter
$0 < \alpha < 1$ quantifies how much a tie strength is reinforced
by each interaction, while $0 <\beta < 1$ accounts for the weakening of the strength of the ties with other individuals.
$w_{max} > 0$ represents the maximum possible value of the weights, which we set to $w_{max} = 1$ 
without loss of generality.
These rules ensure that the weights remain bounded between $0$ and $w_{max}$ 
\add{and,} if a tie's weight is 
zero, it remains so unless there is an interaction involving that tie. Moreover, the weights of the ties between individuals who interact often increase towards $w_{max}$. 
\add{The weights thus represent a quantification of the strength of the social ties at each time, taking into account the history of interactions as well as the impact
of each interaction of an individual on all its ties, strengthening some and weakening others. Interestingly, in a simple case of random and uncorrelated interactions, the
long-time limit of the weight between two individuals can be shown to correspond to their probability of interaction (see Supplementary material).}

It is important to stress that,
 while instantaneous interactions may be undirected
(no source nor target individuals such as in face-to-face interaction data), the evolution rules
\eqref{equation:rule_update_+}-\eqref{equation:rule_update_-} result in a directed network. 
\add{Upon an interaction between $i$ and $j$, $w_{ij}$ and $w_{ji}$ evolve in the same way; however, when $i$ interacts with other individuals than $j$, $w_{ij}$ decreases
while $w_{ji}$ is not affected. For instance, if $j$ interacts only with $i$ but $i$ interacts with many other individuals, $w_{ji}$ can only increase (upon each interaction with $i$),  
while $w_{ij}$ will increase at each interaction of $i$ and $j$ and decrease at each interaction of $i$ with $k \ne j$: $w_{ji}$ thus becomes larger than $w_{ij}$, reflecting the fact that
$i$ is more important to $j$ than $j$ to $i$.
}
\add{Naturally,}
the evolution rules could easily be modified 
in the case of directed interactions, such as in an exchange of text messages or on online social media: for instance, if $i$ sends a message to $j$, the weights $w_{ij}$ and $w_{ik}$ could be affected more strongly than the weights $w_{ji}$ and $w_{jk}$.
However, this would require the introduction of additional parameters.

\subsection*{Application to empirical data}

\begin{figure}[thb]
\centering
\includegraphics[width=\linewidth]{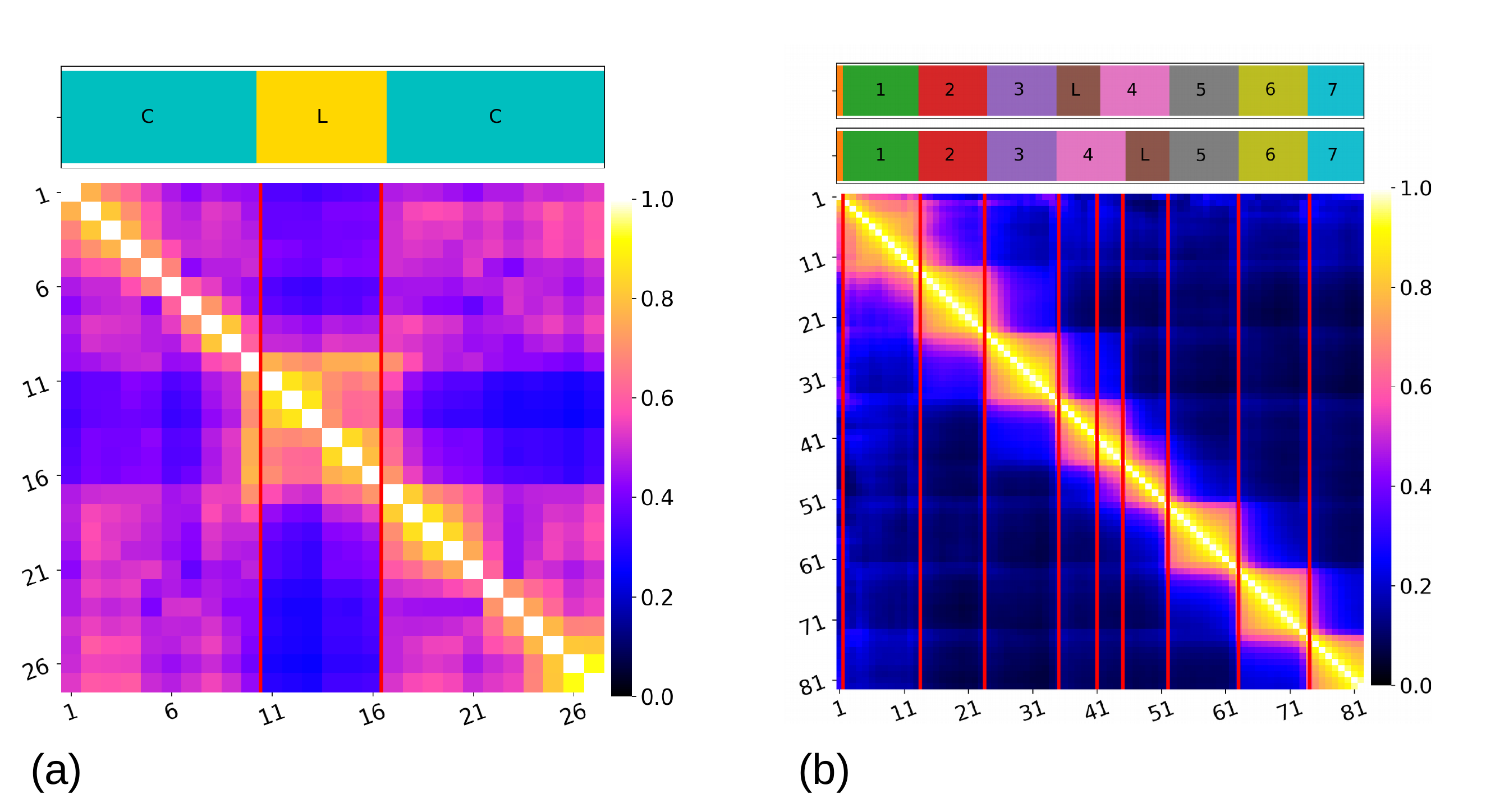}
\caption{ {\bf  Similarity matrices and school schedules
for a day in a French elementary school (a) and in a US middle school (b).}
Here we use $\alpha = \beta = 0.1$, and the evolving networks are observed every $\Delta=20$ minutes for the French school and every $\Delta = 5$ minutes for the US school.
The horizontal bars give information about the schedule of a school day. The different colors correspond to the class times (indicated by the letter C in (a) and with different numbers in (b)) and lunchtimes (indicated by the letter L).
In (b) there are two bars because the students were split into two groups. 
}
\label{fig:gcs_matrices_french_primary_20_middle_utah_5}
\end{figure}

Let us first consider empirical data describing \add{face-to-face proximity} interactions collected by wearable devices in two schools, namely a French elementary school \cite{Stehle:2011} and a US middle school in Utah \cite{Toth:2015role,leecaster2016estimates}, with a temporal resolution of
approximately $20$ seconds in both cases (\add{the devices collected data on the relative proximity of individuals, and not on their location,} see Materials and Methods for more details). Although both cases involve school contexts, the classes were organized very differently, as 
described in \cite{Stehle:2011,leecaster2016estimates}:
the elementary school students remained in the same classroom for their different classes, while the middle school students changed classrooms between classes. 

In each case, we transformed the temporal network data into a network $G(t)$, with the weights
evolving according to the rules \eqref{equation:rule_update_+}-\eqref{equation:rule_update_-}. For simplicity, we used $\alpha=\beta$ and considered various values of $\alpha$. We then stored the network 
 $G(t)$ and the tie weights every 
 $\Delta$ time steps (i.e., we store $G(n \Delta)$ for $n=0,1,2,\cdots$) and computed the similarity between 
each pair of the stored networks
 $G(n \Delta)$ and $G(n' \Delta)$ (see Materials and Methods). We thus obtained a matrix of similarity values \cite{masuda2019detecting,gelardi2019detecting} for each value of $\alpha$, shown in Figure 
\ref{fig:gcs_matrices_french_primary_20_middle_utah_5}
for $\alpha=0.1$.

 \add{For instance, in the case of the US school, the large values of similarity found 
in the diagonal blocks (in yellow) indicate periods in which the network $G(t)$
remains stable, and lower values (off-diagonal) indicate that these periods of stability are different from each other; as seen from the comparison with the school schedule, 
each diagonal block (period of stability of the network) corresponds to a specific class period. In the French school, the organization in blocks correspond to the class
and lunch periods.}
These matrices \add{thus} clearly highlight that the two contexts correspond to different schedules and organizations of interactions
\add{and}
reflect the temporal organization and the  periods of importance in the school schedules.
\add{We show in Figure S1 in the Supplementary material that, 
at small $\alpha$, the weights evolve too slowly and the distinction between the various periods is blurred: the distinction between the various periods becomes 
clearer as $\alpha$ increases}.

\subsection*{Detection of a perturbation}

\begin{figure}[H]
\centering
\includegraphics[width=0.67\linewidth]{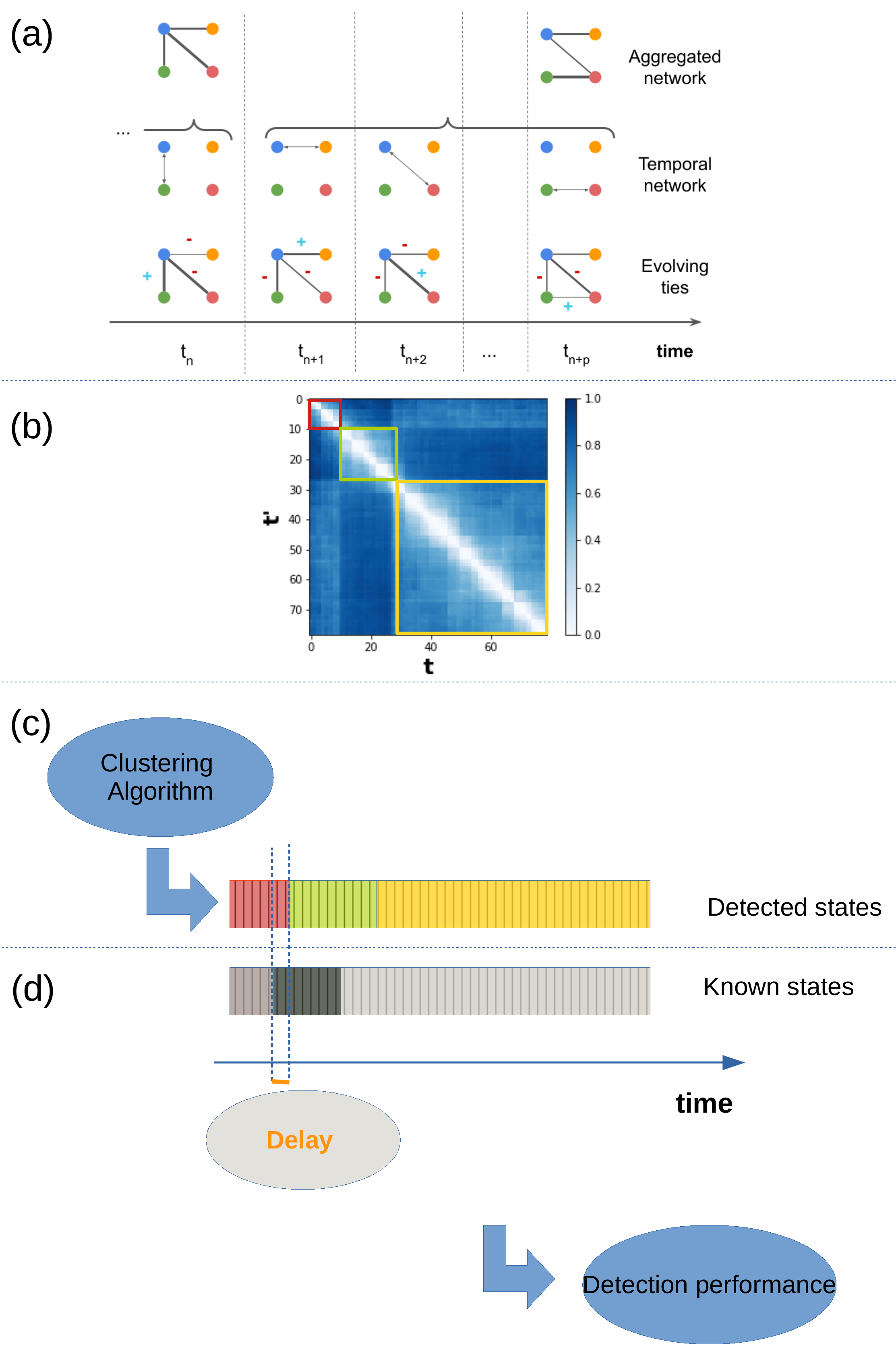}
\caption{{\bf Workflow used to detect discrete states and change points in temporal networks and to estimate the performance of the detection.} 
(a) Creation of a sequence of networks, either by temporal aggregation over successive time windows of $p$ time steps: \add{namely aggregating 
the $p$ snapshots from $t_{mp+1}$ to $t_{mp+p}$ ($m$ being an integer), 
and using as weight $w_{ij}^a$ of a link $ij$ in the aggregated network the number of interactions between $i$ and $j$ in this
time range,}
or by transforming the data into an evolving network observed every $p$ time steps. (b) 
Computation of the similarity between all pairs of networks using the global cosine similarity (see Methods). (c) Classification of the networks into discrete states using a hierarchical clustering algorithm on the distance matrix (the distance between two graphs being simply defined as $1$ minus their similarity). (d) \add{To estimate the performance of the classification, we use (i) the Jaccard index between the known and detected time frame of the perturbation
and (ii) the delay of the detected perturbation, defined as the number of timestamps between
 the actual starting time of the perturbation and the smallest timestamp of the following cluster detected.}
}
\label{fig:detecting_states_flow}
\end{figure}

To go beyond a mere visual inspection of the similarity matrices, we considered a more systematic analysis of the capacity of a temporal network representation, obtained either by temporal aggregation or through our framework, to detect perturbations in a social group's interaction patterns. 
To this aim, we first introduced a synthetic perturbation of controled intensity and duration in the temporal network data, for instance by switching the identity of some nodes for a certain duration. We then followed the steps outlined in Fig.
\ref{fig:detecting_states_flow}. First, 
we used our framework to transform the perturbed temporal network into an evolving weighted graph according to the evolution rules \eqref{equation:rule_update_+}-\eqref{equation:rule_update_-}. This weighted graph was observed every $p$ time steps (if the real time duration of one time step is $\delta$, this means
that we observed the graph every $\Delta= p \delta$). As a baseline, we also aggregated the temporal network
data on successive time windows of duration $\Delta$ (Fig. \ref{fig:detecting_states_flow}a),
\add{i.e., we considered at time $t=(m+1)p$ ($m$ being an integer) the aggregation of the $p$ snapshots $t_{mp+1},t_{mp+2},\dots,t_{mp+p}$, using as 
weights of the aggregated links the number of interactions in that time range.}
We then followed Masuda et al.'s procedure for detecting states in a temporal network \cite{masuda2019detecting}. Namely, 
 we computed the cosine similarity matrix between graphs observed at different times
 (Fig. \ref{fig:detecting_states_flow}b), transformed it
into a distance matrix, and
 applied a hierarchical clustering algorithm
(see Methods) to detect discrete states
of the network.
As the ground truth perturbation is known, 
we added a validation step  to compare the  states obtained by the clustering algorithm to the known perturbation. In this step we quantified the detection performance through two indicators (Fig. \ref{fig:detecting_states_flow}d), namely the Jaccard index between the sets of 
timestamps of the actual perturbation and the timestamps
of the perturbed state detected, and the delay between
the start time of the actual perturbation 
and the corresponding value obtained through the clustering algorithm (see Methods).
  
\begin{figure}[H]
\centering
\includegraphics[width=\linewidth]{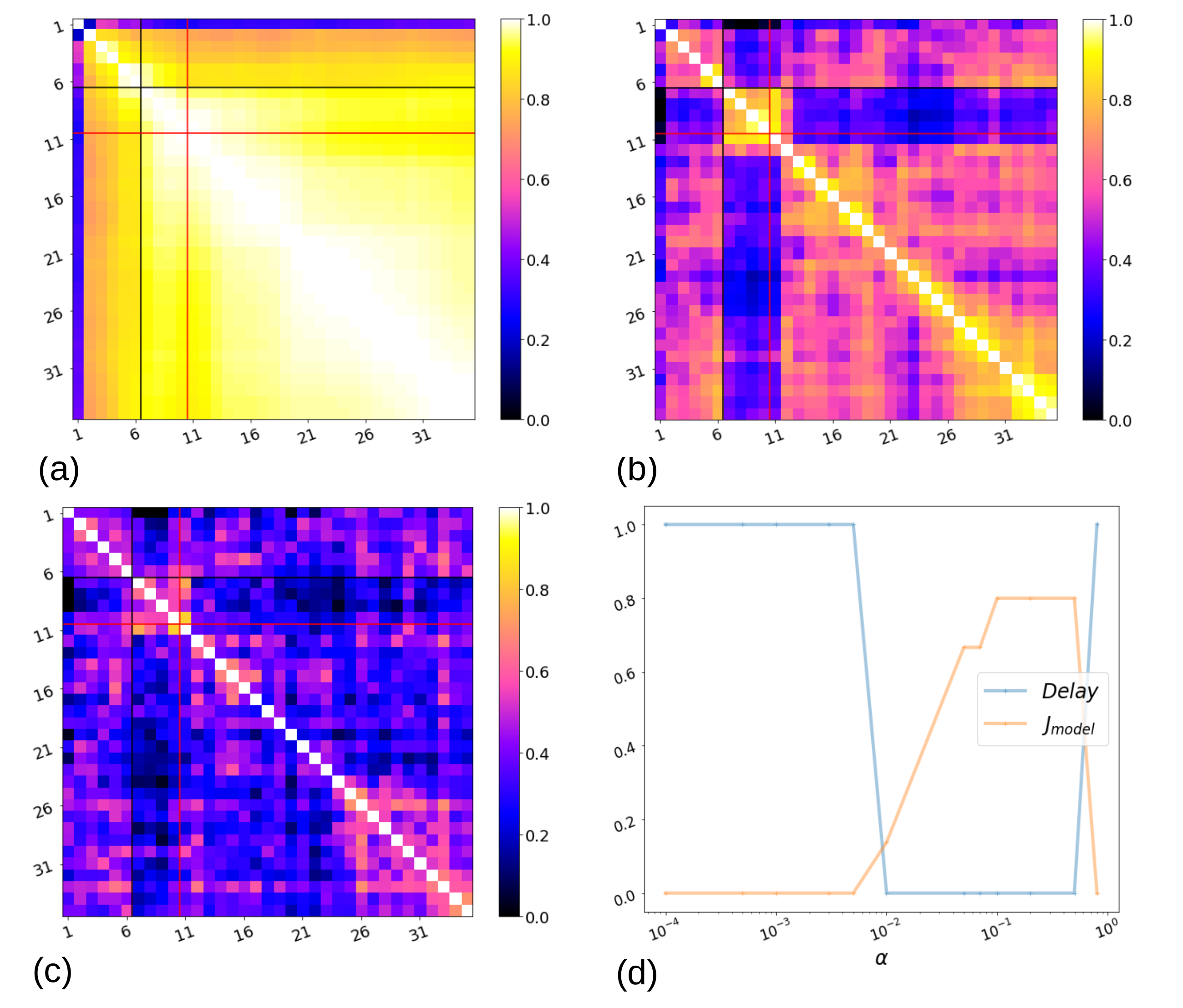}
\caption{{\bf Detection of a simulated perturbation in 
a temporal network data set.} Here we consider one day of proximity data collected from a group of $13$ baboons (see Material and Methods). The data, with a temporal resolution of $20$  seconds, are artificially perturbed by exchanging the identity of two nodes for $2$ hours. The resulting temporal network is transformed into a weighted evolving network as described in the text, and this network is observed every $30$ minutes.
Panels (a), (b), (c) represent the resulting cosine similarity matrices for values of $\alpha = \beta = 0.001$, $0.1$, $0.5$, respectively. The black and red lines correspond to the (known) start and end times of the perturbation. 
Panel (d) shows the performance detection of network states (see Fig. \ref{fig:detecting_states_flow}), computed from the hierarchical clustering analysis applied to the distance matrices, with the number of clusters fixed to $C = 3$. The blue line represents the relative delay in the detection of the perturbation, i.e. the difference between the known beginning of the perturbation (black line) and the detection of a new network state, divided by the total length of the perturbation. The orange line indicates the Jaccard index between the known perturbation
and the perturbation detected by the clustering algorithm. 
The detection performance relative to the aggregated network is not presented because no cluster detected by the algorithm  
\add{corresponded} to the simulated perturbation. \add{The similarity matrices for the aggregated network with different time window lengths are shown in figure S4 of the
Supplementary material.}
}
\label{fig:matrices_jaccard_baboons}
\end{figure}

To illustrate the procedure, we considered proximity data from a group of baboons 
(see Material and Methods). We introduced a small perturbation
in the data, namely the exchange of two individual's identities in the data during a certain period. In Figure
\ref{fig:matrices_jaccard_baboons} we use a perturbation duration of $2$ hours and show the resulting similarity matrices between the weighted evolving networks obtained for three values of $\alpha=\beta$ and observed every $30$ minutes. We also measure and show the detection performance as a function of $\alpha$. Strikingly, even such a small and short perturbation is well detected over a wide range of $\alpha$ values, excepting the smallest and largest.  Notably, the perturbation is instead not detected when using temporal aggregation over successive windows of $30$ minutes. 
The perturbation is not detected for small $\alpha$ values, as the resulting network dynamics is too slow: 
Fig. \ref{fig:matrices_jaccard_baboons}(a) shows that the
network remains very similar to itself during the whole explored time range. However, we observe a sharp increase in detection performance as soon as the resulting dynamics are fast enough. At very large $\alpha$ values, the detection becomes impossible again because each single interaction induces large changes in the weights, leading to rapidly changing dynamics. 

In the supplementary material we also considered daily and weekly time scales of observation with perturbations lasting days or weeks (Figures S5 and S6) 
At such timescales our framework resulted in a perfect or  almost perfect detection of the perturbation for a wide range of values of the parameter $\alpha$ (values of the Jaccard index close or equal to $1$), while the perturbation was rarely detected when using daily aggregated networks.
\add{Note that, at this stage, no single optimal value of the parameter emerges: rather, the existence of the perturbation can be assessed with some
degree of confidence by the fact that the procedure detects the same perturbed state over a range of parameter values.}

\begin{figure}[H]
\centering
\includegraphics[width=\linewidth]{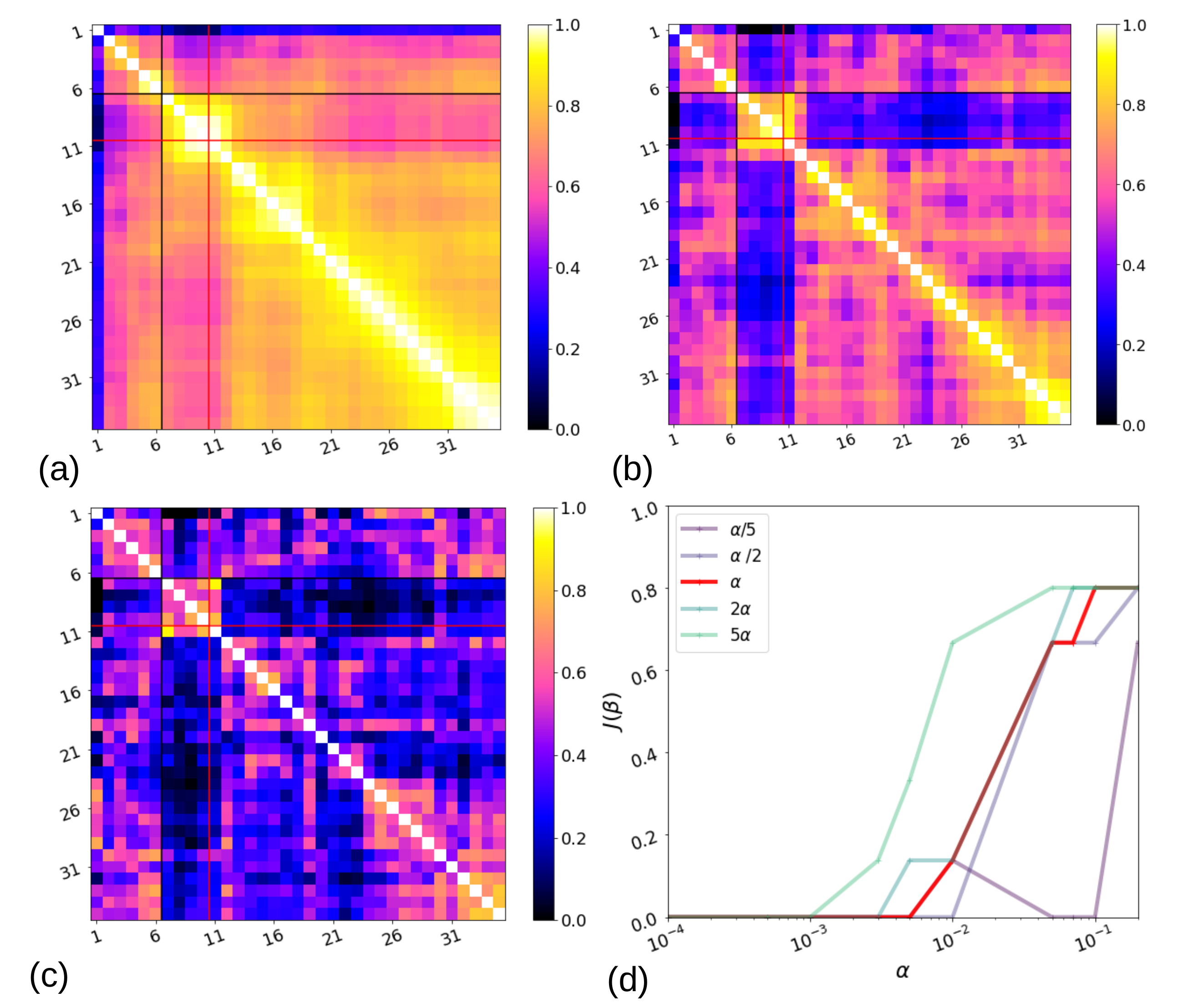}
\caption{{\bf Performance of the detection of simulated changes when varying $\beta$}. Panels (a), (b), (c) represent the cosine similarity matrices  for $\alpha = 0.1$ and values of $\beta = \alpha/5, \alpha, 5\alpha$, respectively, using the same simulated perturbation as in Fig.~\ref{fig:matrices_jaccard_baboons}. Panel (d) shows the performance detection, namely the Jaccard index between the real and detected perturbations, as a function of $\alpha$ and for different values of $\beta$. 
}
\label{fig:jaccard_baboons_2_parameters}
\end{figure}

We further investigated whether using different values for the parameters $\alpha$ and $\beta$ could lead to an improvement in the detection performance. We show the results in 
Figure \ref{fig:jaccard_baboons_2_parameters} for the same data and perturbation as for Figure \ref{fig:matrices_jaccard_baboons}
(see also Supplementary Figure S7). 
We found that the detection performance worsened for $\beta < \alpha$, while it increased for  $\beta > \alpha$.
\add{This can be understood as follows: at small values of $\alpha=\beta$, the weights' increase and decrease are too slow upon a brutal change in the interactions, and the perturbation is not well detected; this can be compensated
by a larger $\beta$ that induces a fast decrease of the weights of non-interacting ties.}
For instance, if a node $i$ was repeatedly interacting with a node $j$ before the perturbation, but interacts more with another one $k$ 
during the perturbation, $w_{ij}$ decreases quickly as soon as the perturbation starts, and this can be easily detected even if 
\add{the small value of $\alpha$ makes} $w_{ik}$ increase only slowly.

\section*{Discussion}

How can we represent the evolution of social relationships? Temporal aggregation procedures  have provided in-depth knowledge on the dynamics of social networks at various timescales \cite{aledavood2015digital,saramaki2014persistence,Fournet:2014,saramaki2014persistence} \add{and} are also used for data-driven numerical simulations of dynamic processes of networks \cite{stehle2011simulation}, 
 possibly with aggregation schemes adapted to the specific process under study \cite{holme2013epidemiologically}. \add{They however lose information on the temporal ordering
 of interactions and do not take into account the impact of interactions on neighbouring ties.}

Here, we have presented a new framework to \add{go from a stream of interactions to a quantification of the strength of ties in a social network and to} study their dynamic evolution, 
based on the idea that social relationships are interdependent: since time resources to invest in social relationships are limited \cite{borgeaud2021vervet}, reinforcing a relationship with someone is necessarily done at the expense of the relationships with others. While 
\add{this idea} can be translated in various ways into specific rules of evolution, here we have focused on a parsimonious two-parameter model rather than on more complex alternatives. We have applied this model to several data sets of interest, showing its ability to highlight changes in the dynamics of the networks and differences between data representing interactions in different contexts. Moreover, we have systematically tested
its ability to detect a perturbation in the network at different timescales. Notably, our results show that this simple model yields a high detection performance 
even for small and short perturbations that cannot be detected by the dynamics of successive aggregated networks. Overall, 
our framework is able to detect perturbations in a broad range of conditions spanning different data sets and various timescales and perturbations. 
This point is particularly important as real-world variations in social relationships can occur on a broad range of timescales, from hours to days to months. For instance, despite decades of research, the timescale of the exchange of favors in primates (e.g., grooming in exchange for other commodities) is still very uncertain  \cite{sanchez2015are}.
Our framework does not require a specification of the timescale of changes to be detected a priori
\add{and in our current study, the temporal organization relevant in the school data sets and the artificial perturbations introduced were detected in a broad range of parameter values. However, whether the same range of parameter values could be used to detect any kind of temporal patterns in an unknown data set remains an open question. To explore a new data set, a sensible path would be either to consider known changes in the network (if such information is available) or to simulate a plausible perturbation of the data, and scan parameter values to check when such changes or perturbations are detected. A natural hypothesis would then be that the temporal patterns present in the data should be detected using the same range of parameter values. Finding the same temporal patterns on a range of parameter values would also give confidence on the significance of these patterns.
}
To investigate this point in more detail, further research will 
use a collection of temporal network models with tunable parameters and different levels of complexity and realism
\cite{perra2012activity,laurent2015from}. Introducing perturbations of various types (e.g., changes in the community structure over time, changes in activity, etc), and of tunable intensity and duration, will allow us to systematically explore the detection capacities and limitations of the evolving weighted graph framework introduced here.

\add{Our focus here has been on the issue of detecting when a perturbation occur in a social network, as it is a known challenging task. Once the existence of a perturbation is established, investigating more detailed quantities such as the distributions of local similarities between the neighborhoods of individuals, or individual trajectories of the similarity of each local neighborhood between successive times (see \cite{gelardi2019detecting}) could also make possible to detect which nodes or links are involved in the perturbation.
Alternatively, this might also be achieved by considering individual matrices of similarities between neighborhoods of individuals at different times, and applying clustering algorithms to each such matrix.}

An interesting property of our framework is that, starting from a stream of undirected interactions, it yields directed ties, because individuals do not invest in their mutual relationship in the same way: for instance, one individual may spend $80\%$ of her time with another, while the other spends only $50\%$ of her time with the first). The weights on each tie can therefore be more or less symmetric, and it would be interesting to investigate 
\add{this point} with respect to the social relationships under study. To this aim, one would need to 
compare the directed network obtained from our framework to other independent measures, such as friendship surveys in a human group or grooming behavior in non-human primates.

While we have limited our current study to a simple version of the model, several extensions could be of interest.
In particular, directed interactions between individuals (such as phone or online messages) could be taken into account, with different impacts on the ties originating from the source of the interaction and on the ties originating from the interaction target. Moreover, one 
could take into account individual characteristics that are often important in relationships by introducing
$\alpha$ and $\beta$ coefficients that depend on individual characteristics such as age, sex, kinship or rank. This would be appropriate for instance when the costs and benefits of interactions differ between low ranking and high ranking individuals \cite{silk1999the}. 
 Our framework \add{could also provide an extension of models of social contagion or consensus formation 
\cite{guilbeault2018complex,rosenthal2015revealing}:
in the spirit of \cite{holme2013epidemiologically}, it could help take into account that interactions with different individuals and at different times are not equivalent, 
by providing a way to dynamically weigh these interactions (an interaction along a currently strong tie would weigh more than along a weak tie).}

Finally, our focus here has been on social relationships of primates in particular, but our conceptual contribution lies in taking into account the interdependence of ties in evolving networks. Thus, our framework may well apply to other systems where such interdependence is relevant, possibly with changes in the rules of evolution.
In particular, we have considered that an interaction between two nodes reinforces the tie between them at the expense of ties with other nodes, but in other contexts, the increase of a tie's weight may in fact increase the importance of related ties. For instance, if a new flight route is created between two airports, passengers may take other flights to connect to other destinations, increasing the traffic on the corresponding routes
\cite{barrat2004weighted}. Taking these interactions into account might open up new perspectives to study the evolution of these types of infrastructure networks \cite{sugishita2020recurrence}.

\section*{Materials and Methods}

{\small
\subsection*{Data Description and Aggregation}

We used three data sets of time-stamped dyadic interactions between individuals corresponding to physical proximity events: 
\begin{itemize}
    \item A data set of contacts between students in an urban public middle school
    in Utah (USA) measured by an infrastructure based on wireless ranging enabled nodes (WRENs) \cite{Toth:2015role,leecaster2016estimates}. The data, available in reference \cite{leecaster2016estimates}, involve 
    679 students in grades 7 and 8 (typical age range from 12 to 14 years old). Participants were recorded over two consecutive days.  
    
    \item A data set gathered by the SocioPatterns collaboration (http://www.sociopatterns.org/) using radio-frequency identification devices in an elementary school in France. These sensors record face-to-face contacts within a distance of about $1.5$m. The data were aggregated with a temporal resolution of $20$ seconds (for more details see \cite{Cattuto:2010}). Contacts between 242 participants (232 elementary school children and 10 teachers) were recorded over two consecutive days \cite{Stehle:2011}. The data are publicly available at http://www.sociopatterns.org/datasets.
    
    \item Data of proximity contacts within a group of Guinea baboons (\emph{Papio papio}), collected from June to November 2019. A subgroup of $13$ baboons consisting only of juveniles and adults were equipped with leather collars fitted with the wearable proximity sensors developed by the SocioPatterns collaboration (see \cite{Gelardi2020MeasuringSN}).

\end{itemize}
}

{\small
\subsection*{Similarity between networks}  

To compare the weighted evolving networks (or aggregated networks) observed at different times, we chose the global cosine similarity between the two vectors formed by the list of all the weights in each network (using a weight $0$ if a link was not present).

A cosine similarity measure is generally defined between two vectors and is bounded between $-1$ and $+1$. It takes the value $+1$ if the vectors are proportional with a positive proportionality constant, a value of $-1$ if the proportionality constant is negative, and $0$ if they are perpendicular. For positive weights, as in our case, it is bounded between $0$ and $1$.

In the case of two networks, $G_1$ and $G_2$, the global cosine similarity is defined as:
\begin{equation}\label{eq:gcs}
    GCS_{G_{1},G_{2}} = \frac{\sum_{i > j} w_{ij}^{(1)}w_{ij}^{(2)}}{\sqrt{\sum_{i > j} \left(w_{ij}^{(1)} \right)^2} \sqrt{\sum_{i > j} \left( w_{ij}^{(2)} \right)^2 }} \ ,
\end{equation}
where the subscripts $^{(1)}$ and $^{(2)}$ denote the weights of the links in the networks $G_1$ and $G_2$, respectively.
}

{\small
\subsection*{Clustering method}  
To obtain discrete system states by hierarchical clustering, we used the "fcluster" function of the 
\verb+scipy.hierarchy+
library from the SciPy module in Python. The function is applied directly on the $t_{max}  \times t_{max}$ distance matrix $d$, obtained by transforming the cosine similarity matrix  elements for each pair of timestamps $(t,t')$: $d(t, t') = 1 - CS(t,t')$. 
To define the distance between clusters, we used the "average" method in the "linkage" function of the library.
We set the number of clusters to $C=3$, corresponding to the periods before, during and after the perturbation.
}

{\small
\subsection*{Detection performance}
To assess the performance of our model, our rationale was that the temporal network representation 
should allow us to detect changes in the social structure, and the quality of the detection entails two aspects: 
it has to be detected (i) without delays and 
(ii) clearly, i.e., social changes have to be distinguished from the noise represented by "ordinary" variations in social activity. In particular, a perturbation is said to be well detected if one of the states found by the clustering algorithm includes all the timestamps of the perturbation and only those.

We first verified that one of the detected clusters could be associated with the perturbation in the data. To this end we determined that each cluster would correspond to a set of contiguous timestamps (thus forming an interval), with the smallest time equal to or larger than the initial timestamp of the perturbation, and largest time equal to or larger than the final timestamp of the perturbation.
A first measure to evaluate the quality of the detection was then given by the "delay" between the actual and the detected perturbation (the number of timestamps between the actual starting time of the perturbation and the smallest  timestamp of the second cluster detected; 
see Figure \ref{fig:detecting_states_flow}d).
 The second measure was given by the Jaccard index $J$ between the set of time steps during which the actual perturbation takes place, ${\cal T}_{ground truth}$, and the set of time steps of the state detected as a perturbation by the clustering procedure,
 ${\cal T}_{detected}$: 
\begin{equation}
    J = \frac{\vert{{\cal T}_{ground truth}} 
    \cap {{\cal T}_{detected}} \vert}
    {\vert {{\cal T}_{ground truth}} \cup {{\cal T}_{detected}} \vert}
\end{equation}

}

\section*{Acknowledgments}{Many thanks to Yousri Marzouki for planting the seed of the idea for this article and to Clément Sire for interesting discussions and the suggestion of studying case $\beta \ne \alpha$ in the model.
A.B. was supported by the ANR project DATAREDUX (ANR-19-CE46-0008) and JSPS KAKENHI Grant Number JP 20H04288.}

\nolinenumbers
\bibliography{biblio}
\bibliographystyle{apalike}

\newpage

\section*{Supporting Information}
\setcounter{section}{0}
\setcounter{table}{0}
\setcounter{equation}{0}
\setcounter{figure}{0}

\renewcommand{\thetable}{S\arabic{table}}
\renewcommand{\thefigure}{S\arabic{figure}}
\renewcommand{\thesection}{Supplementary Note \arabic{section}.}

\section{Interpretation of the weights in a simple case}

We consider the simple example of $N$ nodes among which interactions occur independently and at random, with uniform probability $p$ at each timestep.
We denote by $w_{e}(t)$ the weight of a directed edge $e=(i,j)$. 
At each time step, this weight can either remain the same (if $i$ does not interact with any other node), increase (if $i$ and $j$ interact) or decrease (if $i$ and $j$ do not interact together but
$i$ interacts with $k \ne j$). Thus:
$w_{e}(t+1)=$
\begin{itemize}
    \item[(i)] $(1-\alpha)w_{e}(t)+\alpha$ with probability $p$;
    \item[(ii)] $w_{e}(t)$ with probability $P= (1-p)^{N-1}$;
    \item[(iii)] $(1-\alpha)w_{e}(t)$ with probability $1- p - P$.
\end{itemize}

It is possible to rewrite this recursion relation under the form :
$$w_{e}(t+1)-w_{e}(t)=\alpha \left(-w_{e}(t)+\epsilon_{e}(t)\right) \eta_{e}(t)$$ 
where $\eta_{e}(t)$ and $\epsilon_{e}(t)$ are independent Bernoulli variables of respective parameters $1-P$ and $\frac{p}{1-P}$.
Indeed,
the case $\eta_{e}(t)=1,\epsilon_{e}(t)=1$ occurs with probability $p$ and corresponds to (i); 
the cases $\eta_{e}(t)=0$ correspond to (ii) (and happen with probability $P$); 
the case $\eta_{e}(t)=1,\epsilon_{e}(t)=0$ occurs with probability $(1-P)(1-p/(1-P))=1-p-P$ and
corresponds to (iii).

As $w_{e}(t)$ changes only when $\eta_{e}(t)=1$, we denote by 
$\left(t_{k}\right)_{k\in\mathbb{N}}$ the sequence of times $t_{k}<t_{k+1}$  such that $\eta_{e}(t)\neq0$,  and we 
define $\Tilde{w}_{e}(k)=w_{e}(t_{k})$ and $\Tilde{\epsilon}_{e}(k)=\epsilon_{e}(t_{k})$. Then we have :
\begin{equation}
\Tilde{w}_{e}(k+1)-\Tilde{w}_{e}(k)=\alpha\left(-\Tilde{w}_{e}(k)+\Tilde{\epsilon}_{e}(k)\right)
\label{eq:wek}
\end{equation}
As $t\rightarrow\infty$, the relation between $w$ and $\Tilde{w}$ reads $w_{e}(t)\simeq\Tilde{w}_{e}(t(1-P))$ (as the probability for $\eta$ to be $1$ is $1-P$), 
so in particular they have the same limit at large times. 
We can thus restrict ourselves to the study of $\Tilde{w}_{e}$. 
We can obtain he expression for $\Tilde{w}_{e}(k)$ by writing  \eqref{eq:wek} for $\ell = k-1,k-2,\dots,0$, multiplying each equation by $(1-\alpha)^{k-\ell}$ and summing them. We
then obtain
$$\Tilde{w}_{e}(k)=(1-\alpha)^{k}w_{e}(0)+\frac{\alpha}{1-\alpha}\sum_{l=0}^{k-1}\Tilde{\epsilon}_{e}(l)(1-\alpha)^{k-l},$$
where $w_{e}(0)$ is the initial condition for the weight $w_{e}$.

As the $\Tilde{\epsilon}_{e}(l)$ are identically distributed variables of average $\frac{p}{1-P}$, and using 
$\sum_{l=0}^{\infty}(1-\alpha)^{l}=\frac{1}{\alpha}$, we obtain for the long time limit of the average of $\Tilde{w}_{e}$,
$\mu_{e}=\lim_{t\rightarrow\infty}\left<w_{e}(t)\right>=\lim_{k\rightarrow\infty}\left<\Tilde{w}_{e}(k)\right>$:
$$\mu_{e}=\mu_{e}(\epsilon)=\frac{p}{1-P}$$.

For $p=d/N$ with $d$ finite for $N \to \infty$, i.e., with each node interacting with a finite number of other nodes in the thermodynamic limit at each time, 
$P \sim e^{-d}$ and $\mu_{e}=\frac{p}{1- e^{-d}}$.

As $\mu_{e}\propto p$, we can interpret the long-time limit of the edge weight $w_{i,j}(t)$ as the probability that $i$ interacts with $j$ at time $t$, 
modulated by a factor depending on the properties of the neighbourhood of $i$.

\section{Supplementary figures}

\begin{figure}[H]
\centering
\includegraphics[width=0.9\linewidth]{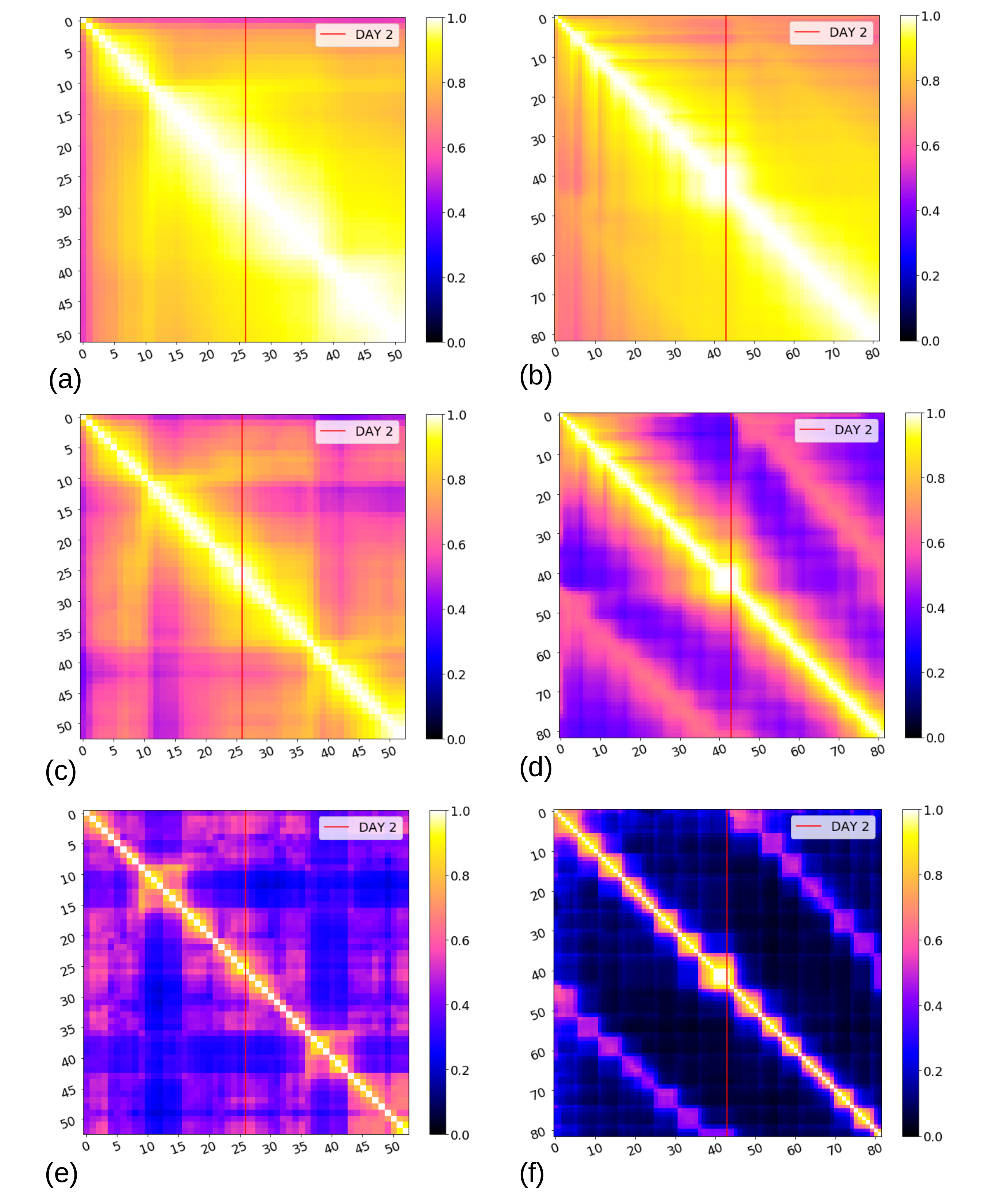}
\caption{ {\bf Cosine Similarity matrices for the French elementary school and Utah middle school, 
using different values of $\alpha$.}
The first column (a),(c),(e) refers to the French elementary school data, sampling the network every 20 minutes; the second column refers to the Utah middle school data, sampling the network every 10 minutes.
The global cosine similarity is calculated between every pair of observed networks. The vertical red lines indicate the time between the first and the second day of data collection (the data of the two days are concatenated).
The different rows refer to three different values of the parameter $\alpha$, namely $\alpha = 0.001$ (a) (b), $\alpha = 0.01$ (c) (d) and $\alpha = 0.1$ (e)(f).
For $\alpha = 0.1$ the different structures of the schools schedules emerge more clearly. We observe for the French school that the structure of the networks during the two lunches are different, and that the structure during the classes remain similar in the two days.
For the Utah middle school, we observe some similarity between the corresponding class periods in the two different days, indicating that the seating arrangements in each class are probably similar in different days. }
\label{fig:gcs_matrices_french_primary_middle_utah_diff_alphas}
\end{figure}

\begin{figure}[h]
\centering
\includegraphics[width=\linewidth]{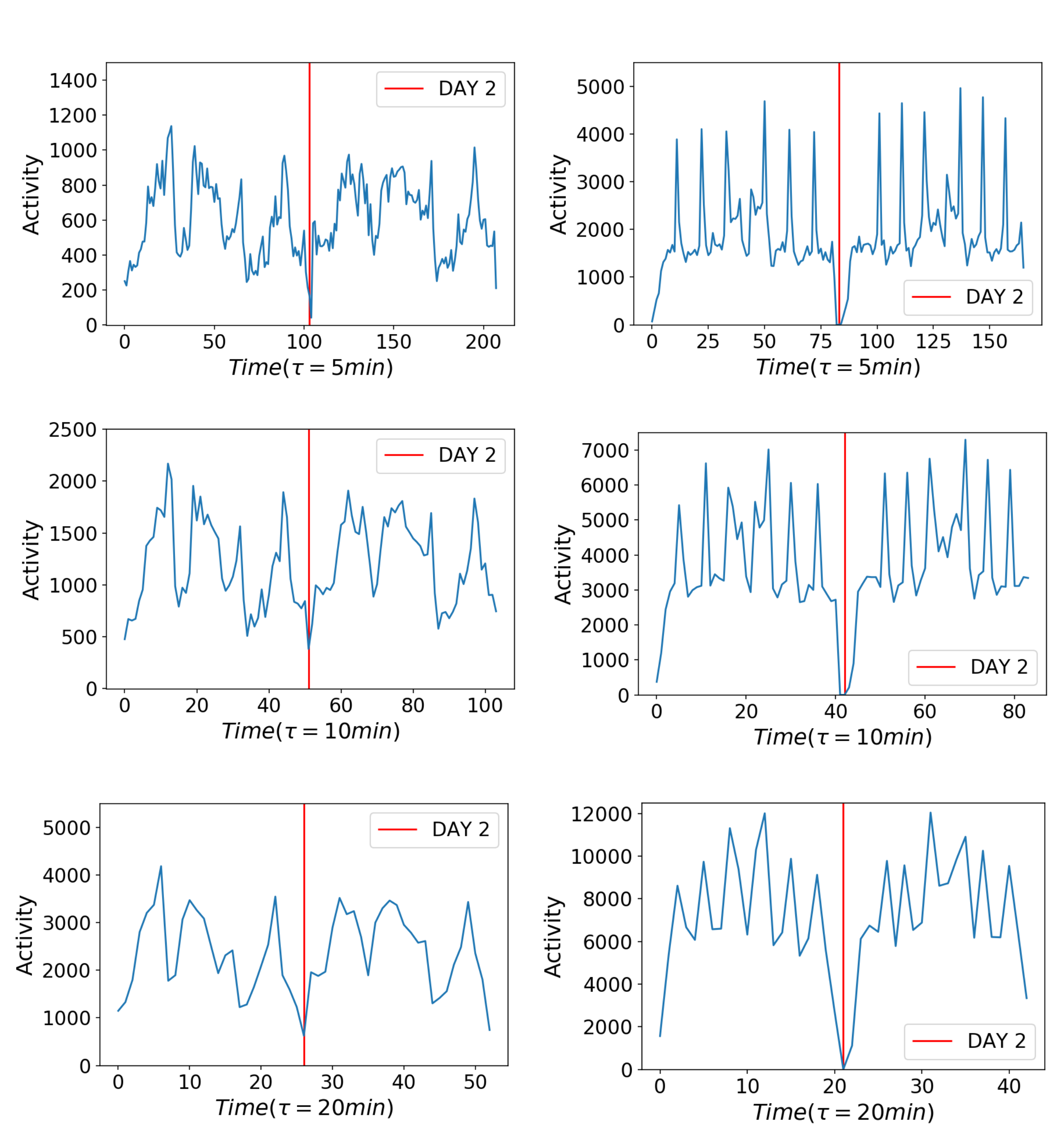}
\caption{{ \bf Event rate (number of temporal edges per timestamp) vs. time} for the French elementary school data (left column) and the Utah middle school data (right column). 
Each row corresponds to a different time resolution (respectively $5$, $10$ and $20$ minutes). While the activity timelines can highlight some timestamps of relevance, they do not shed information on the evolution of the structure of the system (i.e., whether the networks at different times have similar or different structures).
}
\label{fig:activity_vs_time}
\end{figure}

\begin{figure}[h]
\centering
\includegraphics[width=\linewidth]{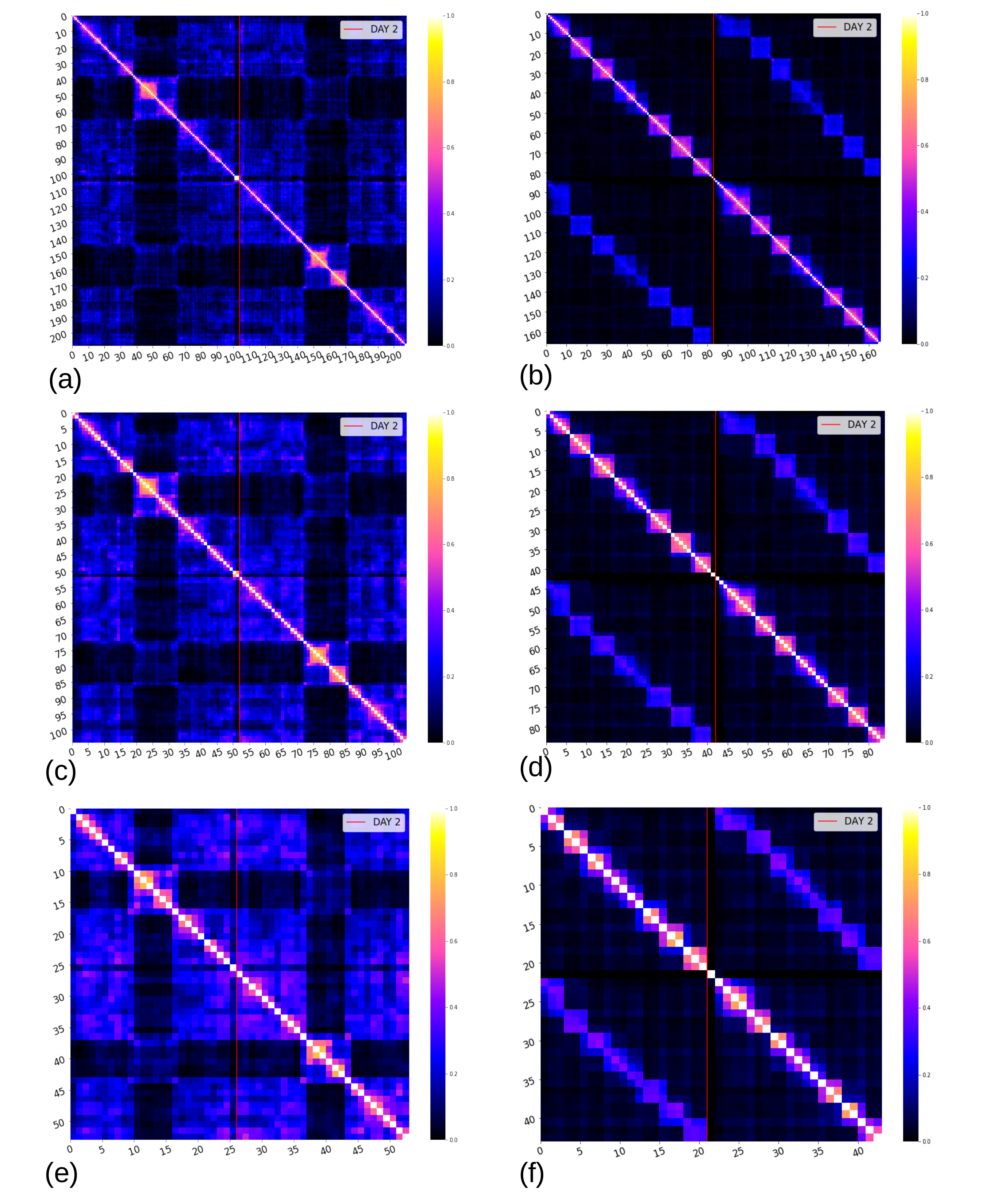}
\caption{ {\bf Cosine Similarity matrices for the French elementary school and Utah middle school, aggregating the network over different time window lengths.}
The first column (a),(c),(e) refers to the French elementary school data ; the second column refers to the Utah middle school data. The different rows refer to three different values of the aggregation time window length: respectively $5$, $10$ and $20$ minutes.}
\label{fig:gcs_matrices_french_primary_middle_utah_aggregated_diff_times}
\end{figure}

\begin{figure}
\centering
\includegraphics[width=0.7\linewidth]{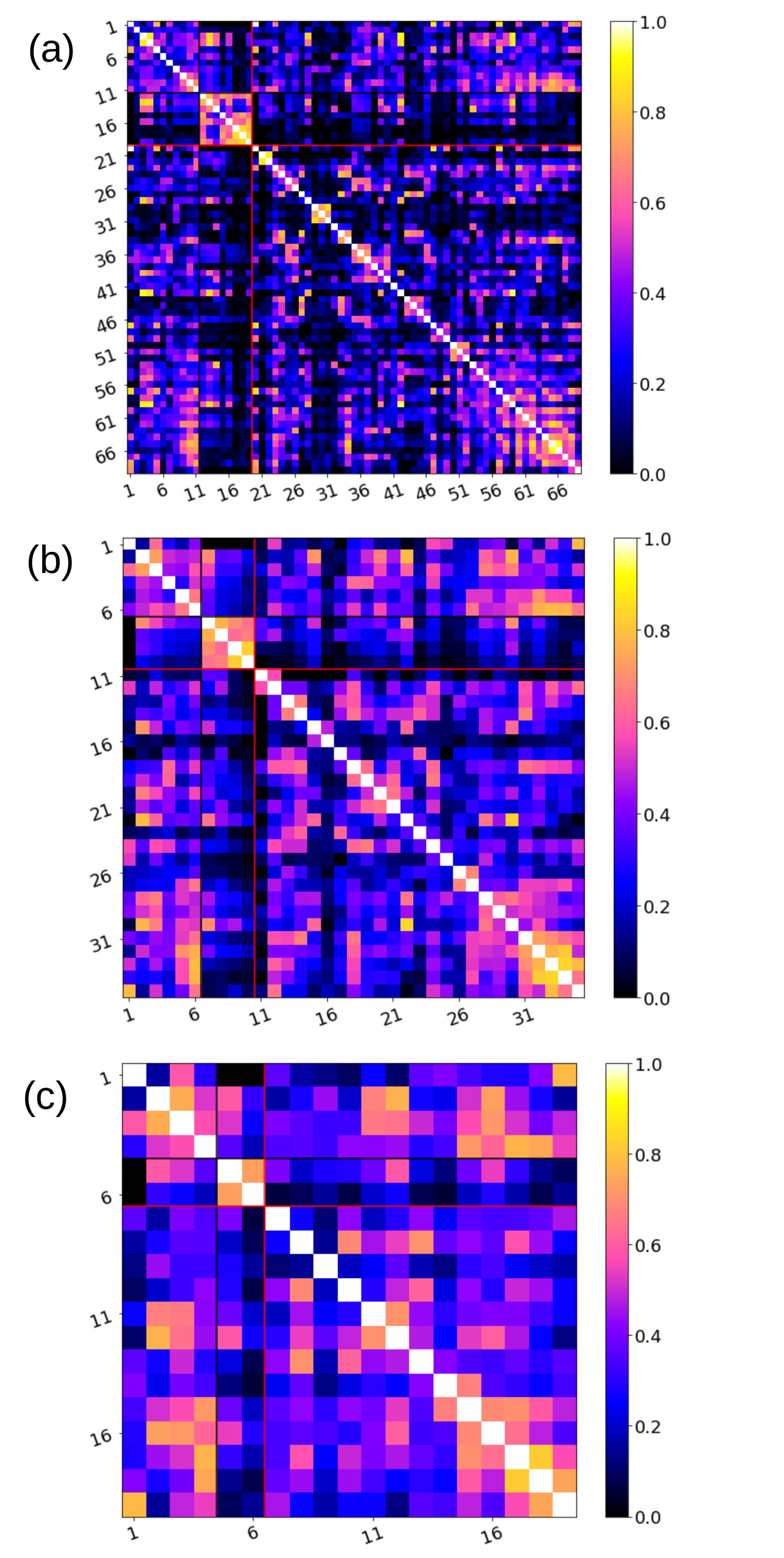}
\caption{ {\bf Cosine Similarity matrices for the baboon data, aggregating the network over different time scales.}
The three panels refer to the aggregated networks computed using the same data as in Fig.~\ref{fig:matrices_jaccard_baboons} for three different values of the aggregation time window, i.e. $15$ minutes (a), $30$ minutes (b) and $1$ hour (c).}
\label{fig:gcs_matrices_baboons_aggregated_diff_times}
\end{figure}

\begin{figure}[thb]
\centering
\includegraphics[width=\linewidth]{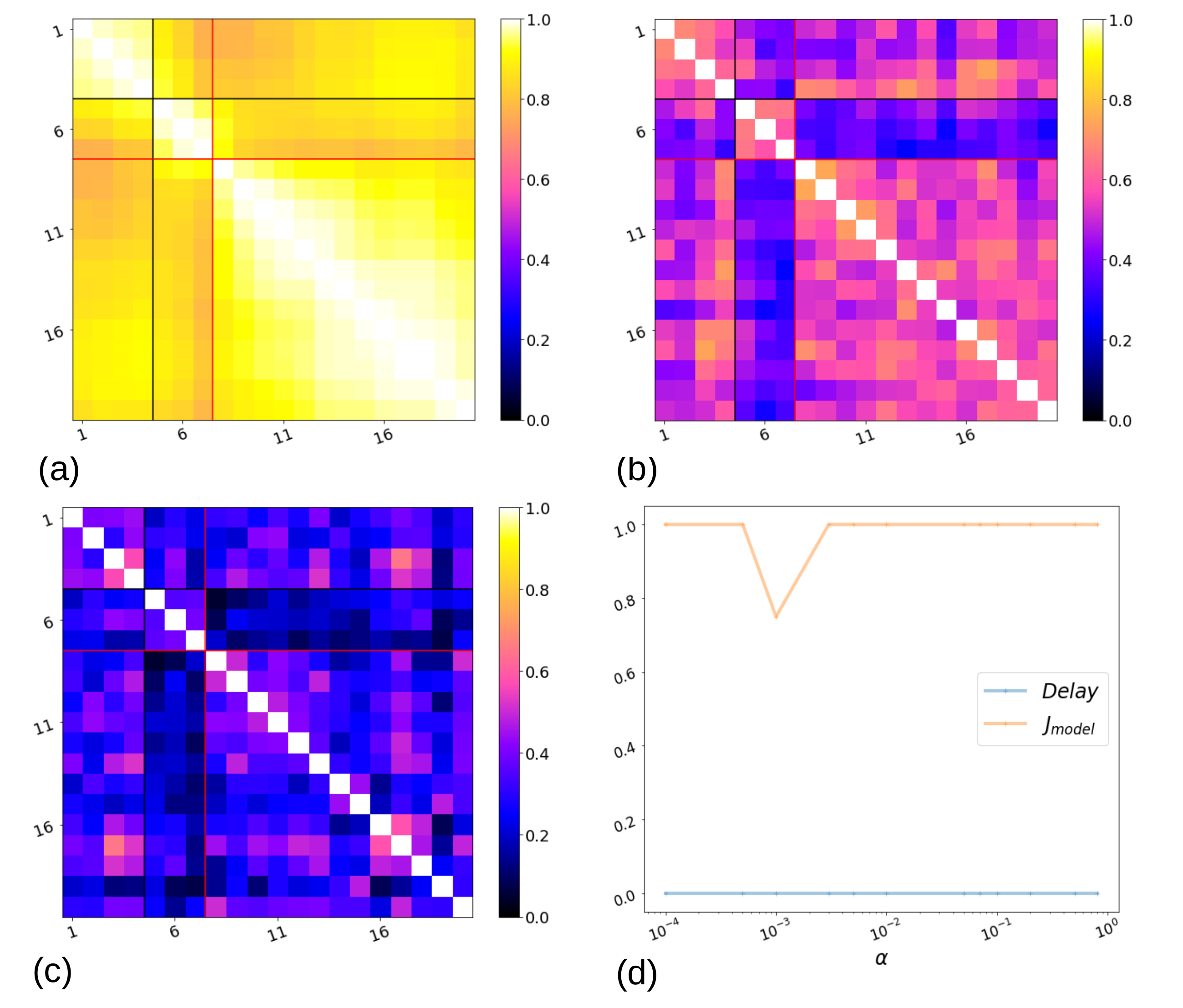}
\caption{Detection of a simulated perturbation in 
a temporal network data set. Here we consider $20$ days of proximity data collected in a group of $13$ baboons (see Material and Methods). The data, whose temporal resolution is $20$ seconds, is artificially perturbed by exchanging the identity of two nodes for $3$ days. The resulting perturbed temporal network is transformed into a weighted evolving network as described in the text, and this network is observed on a daily basis.
Panels (a), (b), (c) represent the resulting cosine similarity matrices for values of $\alpha = \beta = 0.001$, $0.1$, $0.5$, respectively. The black and red lines correspond to the (known) start and end times of the perturbation. 
Panel (d) shows the detection performance (see Fig. \ref{fig:detecting_states_flow}), computed from the hierarchical clustering analysis applied to the distance matrices, with the number of clusters fixed to $C = 3$. The blue line represents the relative delay in the detection of the perturbation, i.e. the difference between the known beginning of the perturbation (black line) and the detection of a new network state, divided by the total length of the perturbation. The orange line indicates the Jaccard index between the known perturbation timestamps
and the perturbation detected by the clustering algorithm using different values of the parameter $\alpha$. 
The detection performance relative to the aggregated network is not shown because no cluster detected by the algorithm could correspond to the simulated perturbation. 
}
\label{fig:matrices_jaccard_baboons_1_day}
\end{figure}

\begin{figure}[thb]
\centering
\includegraphics[width=\linewidth]{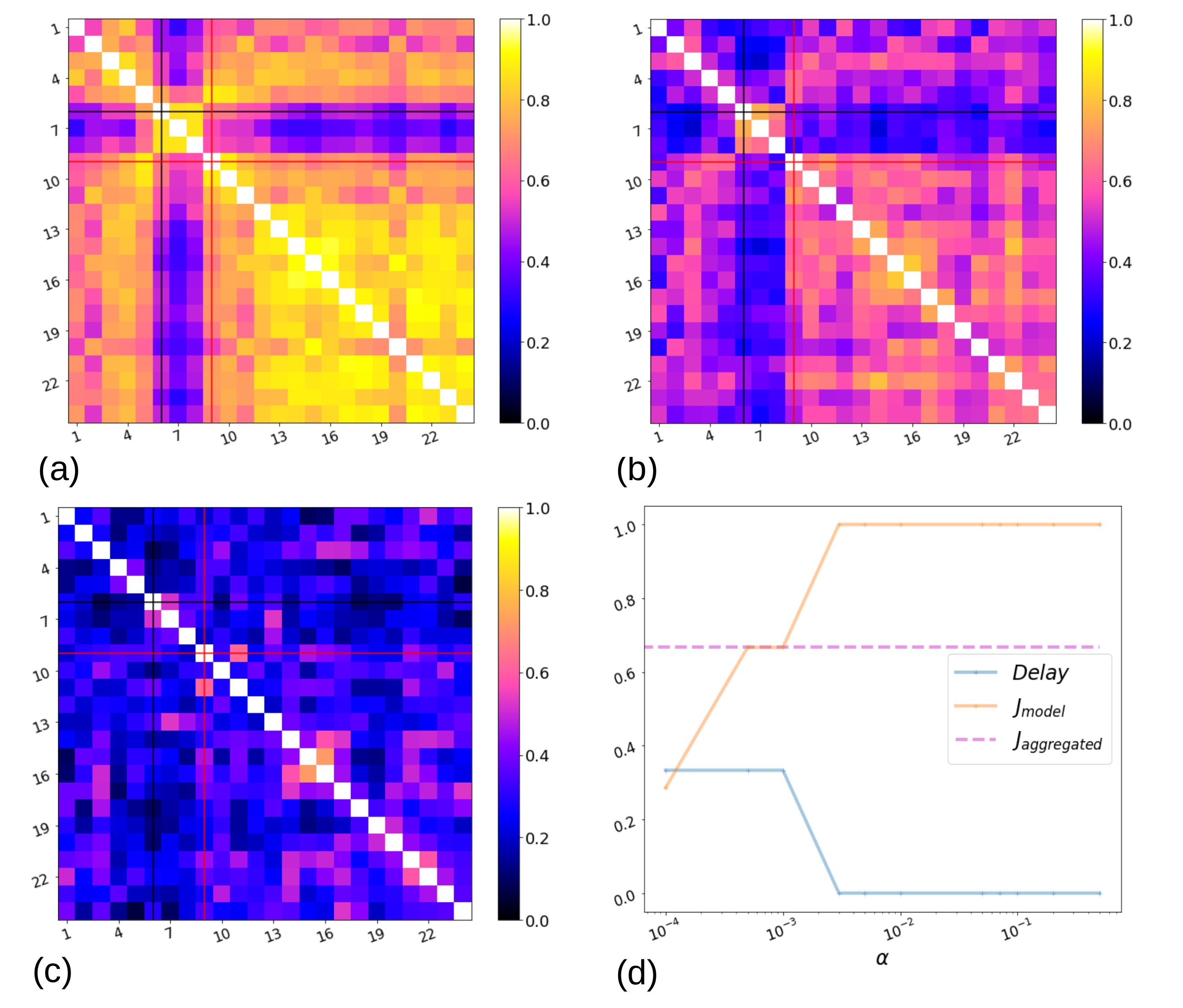}
\caption{Detection of a simulated perturbation in 
a temporal network data set. Here we consider the proximity data collected in a group of $13$ baboons over several months (see Material and Methods). The data, whose temporal resolution is $20$ seconds, is artificially perturbed by exchanging the identity of two nodes during $15$ days, affecting weeks $6$ to $8$. The resulting perturbed temporal network is transformed into a weighted evolving network as described in the text, and this network is here observed every $7$ days.
Panels (a), (b), (c) represent the resulting cosine similarity matrices for values of $\alpha = \beta = 0.001$, $0.1$, $0.5$, respectively. The black and red lines correspond to the (known) start and end times of the perturbation. 
Panel (d) shows the detection performance (see Fig. \ref{fig:detecting_states_flow}), computed from the hierarchical clustering analysis applied to the distance matrices, with the number of clusters fixed to $C = 3$. The blue line represents the relative delay in the detection of the perturbation, i.e. the difference between the known beginning of the perturbation (black line) and the detection of a new network state, divided by the total length of the perturbation. The orange line indicates the Jaccard index between the known perturbation timestamps
and the perturbation detected by the clustering algorithm using different values of the parameter $\alpha$; the dotted magenta line indicates the Jaccard index using networks aggregated over successive $7$ day time windows. 
}
\label{fig:matrices_jaccard_baboons_7_days}
\end{figure}

\begin{figure}[thb]
\centering
\includegraphics[width=\linewidth]{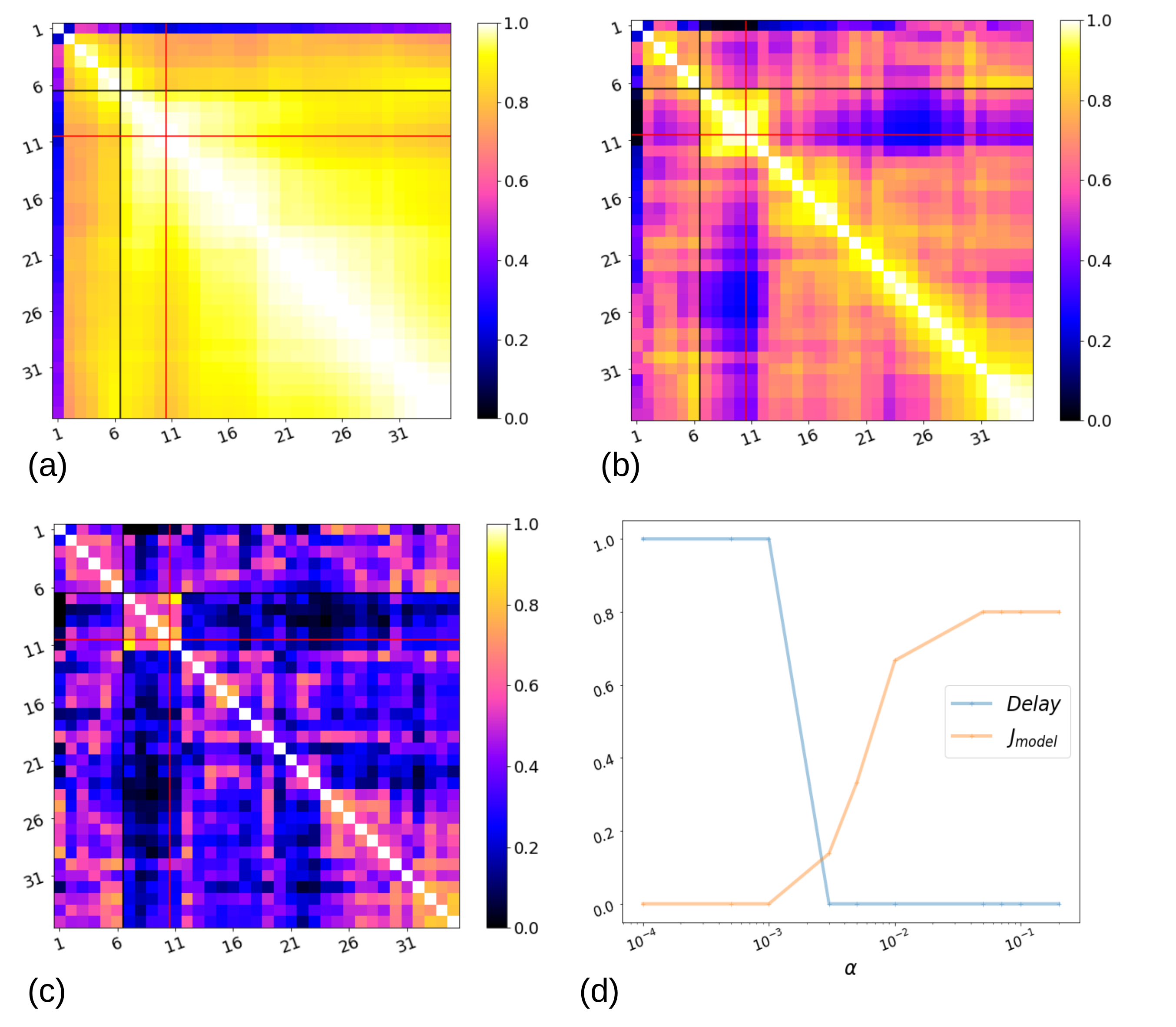}
\caption{{\bf Detection of simulated perturbations in the baboon' data set, using the weighted evolving network with different values of the parameters $\alpha$ and $\beta$.}
Panels (a), (b), (c) represent the cosine similarity matrices for $\beta = 5\alpha$ and values of $\alpha = 0.001, 0.01$ and $0.1$, respectively. The networks and the detection performance were computed using the same procedure as in Fig.~\ref{fig:matrices_jaccard_baboons}. 
}
\label{fig_matrices_jaccard_baboons_2_param}
\end{figure}

\end{document}